\documentclass[11pt, letter]{article}
\pdfoutput=1
\usepackage{sober}
\usepackage{amsthm,amsmath,fullpage,color,graphicx,subfigure,geometry,boxedminipage}
\usepackage{hyperref,verbatim,algorithmic}
\usepackage[section,boxed]{algorithm}

\title{The Forgiving Graph: A distributed data structure for low stretch under adversarial attack} 
\author{Tom Hayes \thanks{Department of Computer Science,  
University of New Mexico,  Albuquerque,  NM 87131-1386;
email: {\tt \{hayes, saia, amitabh\}@cs.unm.edu}. 
This research was partially supported by NSF CAREER Award 0644058,
NSF CCR-0313160, and an AFOSR MURI grant.}
\and Jared Saia   \footnotemark[1]
\and Amitabh Trehan  \footnotemark[1] }

\newtheorem{lemma}{Lemma}

\newtheorem{theorem}{Theorem}

\DeclareGraphicsExtensions{.pdf}

\newcommand{\RT}{\mathrm{RT}}
\newcommand{\RTfragment}{\mathrm{RTfragment}}

\newcommand{\RTparent}{\mathrm{RTparent}}

\newcommand{\Nset}{\mathrm{Nset}}
\newcommand{\hparent}{\mathrm{hparent}}
\newcommand{\hleftchild}{\mathrm{hleftchild}}
\newcommand{\hrightchild}{\mathrm{hrightchild}}
\newcommand{\Endpoint}{\mathrm{endpoint}}
\newcommand{\BT}{\mathrm{BT}}
\newcommand{\PrRoots}{\mathrm{PrRoots}}

\newcommand{\representative}{\mathrm{Representative}}
\newcommand{\edge}{\mathrm{edge}}

\newcommand{\childrencount}{\mathrm{childrencount}}
\newcommand{\height}{\mathrm{height}}
\newcommand{\False}{\mathrm{FALSE}}
\newcommand{\True}{\mathrm{TRUE}}

\newcommand{\haft}{\mathrm{haft}}

\newcommand{\Strip}{\mathrm{Strip}}
\newcommand{\Merge}{\mathrm{Merge}}

\newcommand{\real}{\mathrm{real}}

\newcommand{\numchildren}{\mathrm{numchildren}}
\newcommand{\Haftmergeprint}{\mathrm{HaftMergePrint}}
\newcommand{\size}{\mathrm{size}}

\newcommand{\helper}{\mathrm{helper}}

\newcommand{\hchildren}{\mathrm{hchildren}}

\newcommand{\degree}{\mathrm{degree}}

\renewcommand{\algorithmiccomment}[1]{// #1}

\begin{document}
\date{}
\maketitle

\thispagestyle{empty}

\begin{abstract}
We consider the problem of self-healing in peer-to-peer networks that are under repeated attack by an omniscient
adversary. We assume that, over a sequence of rounds, an adversary either inserts a node with arbitrary connections or
deletes an arbitrary node from the network. The network responds to each such change by quick ``repairs," which consist
of adding or deleting a small number of edges.

These repairs essentially preserve closeness of nodes after adversarial deletions, without increasing node degrees by
too much, in the following sense.   At any point in the algorithm, nodes $v$ and $w$ whose distance would have been $\ell$ in the graph formed by considering only the adversarial insertions (not the adversarial deletions), will be at
distance at most $\ell \log n$ in the actual graph, where $n$ is the total number of vertices seen so far. Similarly, at any
point, a node $v$ whose degree would have been $d$ in the graph with adversarial insertions only, will have degree at most
$3d$ in the actual graph.  Our algorithm is completely distributed and has low latency and bandwidth requirements.

\end{abstract}

\section{Introduction}

Many modern networks are \emph{reconfigurable}, in the sense that the topology of the network can be changed by the
nodes in the network.  For example, peer-to-peer, wireless and mobile networks are reconfigurable.  More generally, many social networks, such as a company's organizational chart; infrastructure networks, such as an airline's transportation network; and biological networks, such as the human brain, are also reconfigurable.  Reconfigurable networks offer the promise of ``self-healing'' in the sense that when nodes in the network fail, the remaining nodes can reconfigure their links to overcome this failure.  In this paper, we describe a distributed data structure for maintaining invariants in a reconfigurable network.  We note that our approach is \emph{responsive} in the sense that it responds to an attack by changing the network topology.  Thus, it is orthogonal and complementary to traditional non-responsive techniques for ensuring network robustness.


This paper builds significantly on results achieved in~\cite{HayesPODC08}, which presented a responsive, distributed data structure called the \emph{Forgiving Tree} for maintaining a reconfigurable network in the fact of attack. The Forgiving Tree ensured two invariants: 1) the diameter of the network never increased by more than a multiplicative factor of $O(\log \Delta)$ where $\Delta$ is the maximum degree in the graph; and 2) the degree of a node never increased by more than an additive factor of $3$.  

In this paper, we present a new, improved distributed data structure called the \emph{Forgiving Graph}.  The improvements of the Forgiving Graph over the Forgiving Tree are threefold.  First, the Forgiving Graph maintains 
low stretch i.e. it ensures that the distance between any pair of nodes $v$ and $w$ is close to what their distance would be even if there were no node deletions.  It ensures this property even while keeping the degree increase of all nodes no more than a multiplicative factor of $3$.  Moreover, we show that this tradeoff between stretch and degree increase is asymptotically optimal.  Second, the Forgiving Graph handles both adversarial insertions and deletions, while the Forgiving Tree could only handle adversarial deletions (and no type of insertion).  Finally, the Forgiving Graph does not require an initialization phase, while the Forgiving Tree required an initialization phase which involved sending $O(n \log n)$ messages, where $n$ was the number of nodes initially in the network, and had a latency equal to the initial diameter of the network.  Additionally, the Forgiving Graph is divergent technically from the Forgiving Tree, it makes significant use of a novel distributed data structure that we call a Half-full Tree (HaFT). 

\medskip
\noindent {\bf Our Model:} We now describe our model of attack and
network response, which is identical to that of~\cite{HayesPODC08}.  We assume that the network is initially a connected
graph over $n$ nodes.  An adversary repeatedly attacks the
network. This adversary knows the network topology and our
algorithm, and it has the ability to delete arbitrary nodes from the
 network or insert a new node in the system which it can connect to any subset of the nodes currently in the system.  
However, we assume the adversary is constrained in that in any time step it can only delete or insert a single node. 

\medskip
\noindent {\bf Our Results:}  For a peer-to-peer network that has both insertions and deletions, let $G'$ be the graph consisting of the original nodes and inserted nodes without any changes due to deletions. Let $n$ be the number of
nodes in $G'$. The Forgiving Graph ensures that: 1) the distance between any two nodes of the actual network never increases by more than $\log n$ times their distance in $G'$; and 2) the degree of any node never increases by more than $3$ times its degree in $G'$.  Our algorithm is completely distributed and resource efficient.  Specifically, after deletion, repair takes
$O(\log d\log n)$ time and requires sending $O(d\log n)$ messages, each of size $O(\log n)$ where $d$ is the degree
of the node that was deleted. The formal statement and proof of these results is in Section~\ref{subsec: upperbounds}. 

\medskip
\noindent {\bf Related Work:} 
Our work significantly builds on work in~\cite{HayesPODC08} as described above.  There have been numerous other papers that
discuss strategies for adding additional capacity or rerouting in
anticipation of failures \cite{ doverspike94capacity,
frisanco97capacity, iraschko98capacity, murakami97comparative,
caenegem97capacity, xiong99restore}.  Results that
are responsive in some sense include the following.  M\'{e}dard, Finn, Barry, and Gallager
\cite{medard99redundant} propose constructing redundant trees to make
backup routes possible when an edge or node is deleted.  Anderson,
Balakrishnan, Kaashoek, and Morris \cite{anderson01RON} modify some
existing nodes to be RON (Resilient Overlay Network) nodes to detect
failures and reroute accordingly. Some networks have enough redundancy
built in so that separate parts of the network can function on their
own in case of an attack~\cite{goel04resilient}.  In all these past
results, the network topology is fixed.  In contrast, our approach
adds edges to the network as node failures occur.  Further, our
approach does not dictate routing paths or specifically require
redundant components to be placed in the network initially.   Our model of attack and repair builds on earlier work in~\cite{BomanSAS06, SaiaTrehanIPDPS08}.


There has also been recent research in the physics community on
preventing cascading failures.  In the model used for these results,
each vertex in the network starts with a fixed capacity. When a vertex
is deleted, some of its ``load'' (typically defined as the number of
shortest paths that go through the vertex) is diverted to the
remaining vertices.  The remaining vertices, in turn, can fail if the
extra load exceeds their capacities. Motter, Lai, Holme, and Kim have
shown empirically that even a single node deletion can cause a
constant fraction of the nodes to fail in a power-law network due to
cascading failures\cite{holme-2002-65, motter-2002-66}. Motter and Lai
propose a strategy for addressing this problem by intentional removal
of certain nodes in the network after a failure begins
\cite{motter-2004-93}.  Hayashi and Miyazaki propose another strategy,
called emergent rewirings, that adds edges to the network after a
failure begins to prevent the failure from
cascading\cite{hayashi2005}.  Both of these approaches are
shown to work well empirically on many networks.  However, unfortunately, they
perform very poorly under adversarial attack.

\section{Node Insert, Delete and Network Repair Model}
\label{sec:prelim}


\begin{figure}[h!]
\caption{The Node Insert, Delete and Network Repair Model -- Distributed View.}
\label{algo:model-2}
\begin{boxedminipage}{\textwidth}
\begin{algorithmic}
\STATE Each node of $G_0$ is a processor.  
\STATE Each processor starts with a list of its neighbors in $G_0$.
\STATE Pre-processing: Processors may send messages to and from
their neighbors.
\FOR {$t := 1$ to $T$}
\STATE Adversary deletes or inserts a node $v_t$ from/into $G_{t-1}$, forming $H_t$.
\IF{node $v_t$ is inserted} 
\STATE The new neighbors of $v_t$ may update their information and send messages to and from
their neighbors.
\ENDIF
\IF{node $v_t$ is deleted} 
\STATE All neighbors of $v_t$ are informed of the deletion.
\STATE {\bf Recovery phase:}
\STATE Nodes of $H_t$ may communicate (asynchronously, in parallel) 
with their immediate neighbors.  These messages are never lost or
corrupted, and may contain the names of other vertices.
\STATE During this phase, each node may insert edges
joining it to any other nodes as desired. 
Nodes may also drop edges from previous rounds if no longer required.
\ENDIF
\STATE At the end of this phase, we call the graph $G_t$.
\ENDFOR
\vspace{10pt}
\hrule
\STATE
\STATE {\bf Success metrics:} Minimize the following ``complexity'' measures:\\
Consider the graph  $G'$ which is the graph consisting solely of the original nodes and insertions without regard to
deletions and healings. Graph $G'_{t}$ is $G'$ at timestep $t$ (i.e. after the $t^{\mathrm{th}}$ insertion or deletion).
  \begin{enumerate}
\item{\bf Degree increase.}  $\max_{v \in G} \degree(v,G_T) / \degree(v,G'_T)$
\item {\bf Network stretch.} $\max_{x, y \in G_{T}} \frac{dist(x,y,G_{T})}{dist(x,y,G'_{T})}$, where, for a graph $G$ and nodes $x$ and $y$ in $G$, $dist(x,y,G)$ is the
length of the shortest path between $x$ and $y$ in $G$.
\item{\bf Communication per node.} The maximum number of bits sent by a single node in a single recovery round.
\item{\bf Recovery time.} The maximum total time for a recovery round,
assuming it takes a message no more than $1$ time unit to traverse any edge and we have unlimited local computational power at each node.
\end{enumerate}
\end{algorithmic}
\end{boxedminipage}
\end{figure}
We now describe the details of our node insert, delete and network repair model.  Let $G = G_0$ be an arbitrary graph on
$n$ nodes,
which represent processors in a distributed network.  In each step, the adversary either deletes or adds a node. 
After each
deletion, the algorithm gets to add some new edges to the graph, as well as deleting old ones.  At each insertion, the
processors follow a protocol to update their information.
The algorithm's goal is to maintain connectivity in the network, keeping the distance between the nodes small.  At the
same time, the algorithm wants to
minimize the resources spent on this task, including keeping node degree small.  


Initially, each processor only knows its neighbors in $G_0$, and is unaware of the structure of the rest of $G_0$.
After each deletion or insertion, only the neighbors of the deleted or inserted vertex are informed that
the deletion or insertion has occured. After this, processors are allowed to communicate by sending a limited number
of messages to their direct  neighbors.  We assume that these messages are always sent and received successfully.  The
processors may also request new edges be added to the graph. The only synchronicity assumption we make is that no
other  vertex is deleted or inserted until the end of this round of computation and communication has concluded.
To make this assumption more reasonable, the per-node communication cost should be very small in $n$ (e.g. at most logarithmic).

We also allow a certain amount of pre-processing to be done before the first attack occurs.  This may, for instance,
be used by the processors to gather some topological information about $G_0$, or perhaps to 
coordinate a strategy.  Another success metric is the amount of computation and communication needed during this
preprocessing round.  Our full model is described in Figure~\ref{algo:model-2}.


\section{The Forgiving Graph algorithm}
\label{sec:algorithm}

\begin{figure}[t!]
\centering
\includegraphics[scale=0.5]{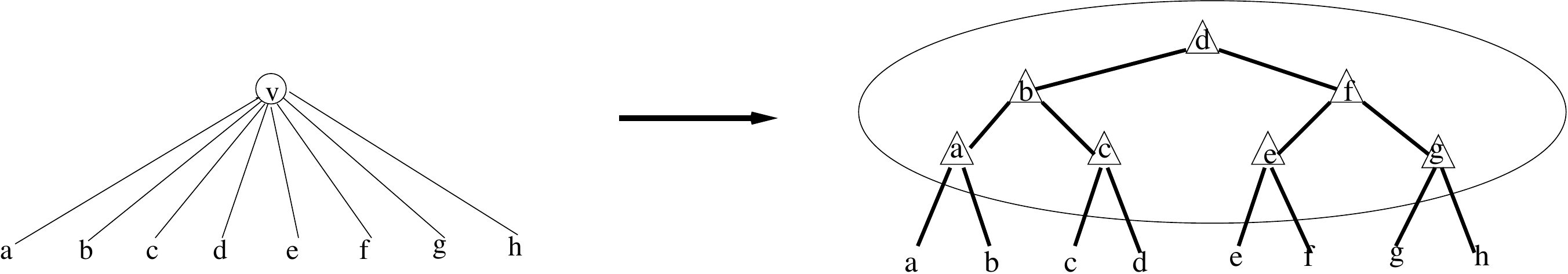}
\caption{Deleted node $v$ replaced by its Reconstruction Tree. The nodes in the triangle are helper nodes
simulated by the real nodes which are in the leaf layer.}
 \label{fig: RT}
\end{figure}

At a high level, our algorithm works as follows:

In our model, an adversary can effect the network in one of two ways: inserting a new node in the network or deleting an
existing node from the network. Node insertion is straightforward and is dependent on the specific policies of the
network. When an insertion happens, our incoming node and its neighbors update the data structures that are
used by our algorithm. We will also assume that nodes maintain neighbor-of-neighbor information.

Each time a node $v$ is deleted, we can think of it as being replaced by a Reconstruction Tree ($\RT(v)$, for short) which
is a  $\haft$ (discussed in  Section~\ref{sec: hafts}) having ``virtual'' nodes as internal nodes and
neighbors of $v$ as the leaf nodes. Note that each virtual node has a degree of at most  $3$. A single real node itself is a trivial $\RT$ with one node.
 $\RT(v)$ is formed by merging all the neighboring $\RT$s of
$v$ using the strip and merge operations from Section~\ref{sec: hafts}.
After a long sequence of such insertions and deletions, we are left with a graph which is a patchwork mix of virtual
nodes and original nodes. 


Also, because the virtual trees (hafts) are balanced binary trees, the deletion of a node $v$ can, at worst, cause the
distances between its neighbors to increase from $2$ to $2 \lceil \log d  \rceil$ by travelling through its $\RT$, where 
$d$ is the degree of $v$ in $G'$ (the graph consisting solely of the original nodes and insertions without regard to
deletions and healings). 
However, since this deletion may cause many $\RT$s to merge and the new $\RT$ formed may
involve all the nodes in the graph, the distances between any pair of actual surviving nodes may increase by no
 more than a $\lceil \log n \rceil$ factor.
 
Since our algorithm is only allowed to add edges and not nodes, 
we cannot really add these virtual nodes to the network.
We get around this by assigning each virtual node to an actual
node, and adding new edges between actual nodes in order to 
allow ``simulation'' of each virtual node.  More precisely,
our actual graph is the homomorphic image of the graph
described above, under a graph homomorphism which fixes 
the actual nodes in the graph and maps each virtual node
to a distinct actual node which is ``simulating'' it.


Note that, because each actual node simulates at most one
virtual node for each of its deleted neighbors, and virtual nodes have degree at most $3$,
this ensures that the maximum degree increase of our algorithm
is at most $3$ times the node's degree in $G'$.

\section{Half-full Trees}
\label{sec: hafts}

\begin{figure}[h!]
\centering
\subfigure[A haft with 7 leaf nodes.]{ \label{sfig: haftexample} \includegraphics{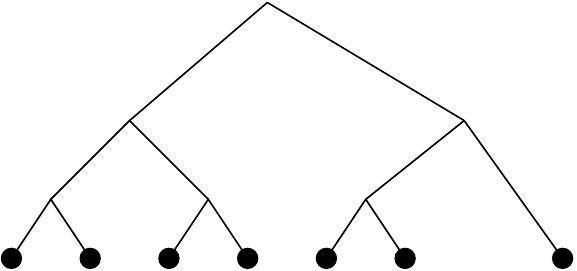} }
\hspace{0.4in}
\subfigure[A haft of n leaves. Every haft is a union of complete binary trees. In our notation, $T_a$ is a complete binary tree and $|T_{a}|$ is the number of leaf nodes in $T_{a}$. The nodes in the square boxes are the nodes not part of a complete tree.]{ \label{sfig: haft-as-join} \includegraphics{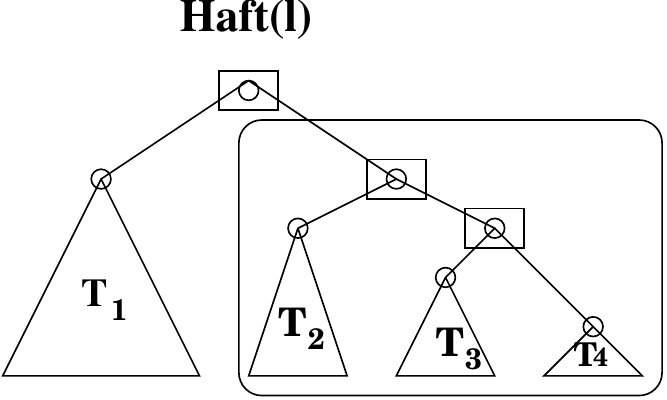} }
\caption{haft (half-full tree)}
\end{figure}

 This section defines half-full trees (\emph{haft}, for short), and describes some of their interesting properties of concern to us.



\begin{description}
\item[Half-full tree:] A haft is a rooted binary tree in which every non-leaf node $v$
has the following properties:
\begin{itemize}
 \item $v$ has exactly two children.
 \item The left child of $v$ is the root of a complete binary subtree, containing  half or more of $v$'s  descendants.
\end{itemize}
\end{description}

 An example of a haft is shown in figure \ref{sfig: haftexample}. For any positive $l$, there is a single unique haft over $l$ leaf nodes (see lemma~\ref{lemma: hftproperties}), that we refer to as as $\haft(l)$

\begin{lemma}
\label{lemma: hftproperties}
Let $l$ be a positive integer. Then, the following are true:
\begin{enumerate}
\item \label{lmitem: hftprop-unique} There is a single unique $\haft$ with $l$ leaf nodes, that we refer to as $\haft(l)$.
\item \label{lmitem: hftprop-bin} binary representation (one-to-one correspondence): 
 Let $a_{k}a_{k-1}...a_{0}$ be the binary representation of $n$. Let $h$ be the
number of ones  in this representation. Let $x_{1}, x_{2}, \ldots, x_{h}$ be the indices of the one bits, and $n =
\sum_{i=1}^{h} 2^{x_{i}}$, sorted in descending order. Let $T_{i}$ be the complete
binary tree with $2^{x_{i}}$ leaves. We can break $\haft(l)$ into a forest of $h$ complete binary trees ($T_{1}, T_{2},
\ldots T_{h}$) by removing $h-1$ nodes from $T$. 
\item \label{lmitem: hftprop-depth} The depth of $\haft(l)$ is $\lceil \log l \rceil$.
\end{enumerate}
\end{lemma}
\begin{proof}
We now prove parts~\ref{lmitem: hftprop-unique} and \ref{lmitem: hftprop-bin}. Let $T$ be a haft on $l$ leaves. As a
running example, consider the  $\haft$ shown in Figure~\ref{sfig: haft-as-join}.   Let $a_{k}a_{k-1}...a_{0}$ be the binary
representation of $l$. Let $h$ be the number of ones  in this representation. Let $x_{1}, x_{2}, \ldots, x_{h}$ be the
indices of the one bits sorted in descending order, and $l = \sum_{i=1}^{h} 2^{x_{i}}$. Let $T_{i}$ be the complete
binary tree with $2^{x_{i}}$ leaves. By definition of a haft, there are two cases:
\begin{enumerate}
\item  \emph{ $T$ is a complete tree}: This happens when $h=1$ and $n = 2^{x_{1}}$. Clearly, $T$ is unique, corresponding
to the complete tree $T_{1}$.
\item \emph{ $T$ is not a complete tree}: By definition of $\haft$, the left child of the root is a complete tree and
moreover this tree has half or more of the children of the root. Let $Size(X)$ be the number of nodes in a tree $X$.
Since $Size(T_{i}) =  2^{x_{i+1}} - 1$ we know that $Size(T_{1}) > \sum_{k=2}^{h} Size(T_{k}) $. Thus, the
complete tree to the left of the root has to be $T_{1}$. \\
 Applying the same definition to the right child of the root, we see that either this node heads the tree $T_{2}$, or
its left subtree is $T_{2}$. Recursively applying this reasoning,  we see that $\haft(l)$ is a unique tree with
the trees $T_{1}$ to $T_{h}$ joined by $h-1$ single nodes (For example in in Figure~\ref{sfig:
haft-as-join}, these $h-1$ single nodes are marked as square boxes ). It directly follows  that removing these $h-1$
nodes leaves us with a forest of $h$ complete binary trees $T_{1}, T_{2},\ldots T_{h}$.
\end{enumerate}

 For part \ref{lmitem: hftprop-depth}, there are two possibilities:
 \begin{enumerate}
 \item \emph{$T$ is a complete tree:} For a complete tree with $l$ leaves, we know that the depth of the tree is $\log
l$.
\item \emph{$T$ is not a complete tree:} 
We show this by induction on the number of leaf nodes. Consider a $\haft$ with $l$ leaf nodes. If $l=1$, the $\haft$ is
a complete tree so the height is $0$, which is $\log l$. For larger $l$, we note that the left child of the root heads
a complete subtree with less than $l$ leaf nodes. Thus, the height of this left subtree is no more than $\log l$.
Moreover, the right child of the root heads a $\haft$ over no more than $\frac{l}{2}$ leaf nodes. Thus, by the
inductive hypothesis, this right subtree has height at most $\lceil \log\frac{l}{2} \rceil$. Thus, the height of
$\haft(l)$ is $1 + max(\log x, \log(l - x))$, where $x$ is a power of 2 and $\frac{l}{2} \le x < n$. Since $x > l- x$,
it follows that $\log x = \lceil log x \rceil \ge \lceil \log(n - x) \rceil$. Finally, the height of $\haft(l)$ is $1 +
\log x = \lceil\log l \rceil$, since $\frac{l}{2} \ge x < l$.

\end{enumerate}

\end{proof}

\subsection{Operations on Hafts}
We Define the following operations on hafts:
\begin{enumerate}
\item \emph{Strip}: Suppose $T$ is a haft with $h$ ones in its binary representation. The Strip operation removes 
$h-1$ nodes from $T$ returning  a forest of $h$ complete trees. 

\item \emph{Merge}: The Merge operation joins  hafts together using additional isolated single nodes,
to create a single new haft.
\end{enumerate}

We now describe these operations in more detail:

\subsubsection{Strip}
\label{subsec: haftstrip}
The operation $\Strip(T)$ takes a $\haft$ $T$ and returns a forest $F,$ of complete trees.  As follows from part \ref{lmitem: hftprop-bin} of lemma \ref{lemma: hftproperties}, each $\haft$ can be broken into a forest of $h$ complete trees where $h$ is the number of ones in the binary representation of the number of leaves of $T$. We call the roots of these complete trees primary roots. Before we proceed further, let us formally define this concept:

\begin{description}
\item{Primary root:} A primary root is a node in a $\haft$ that has the following properties:
\begin{itemize}
\item It is the root of a complete subtree.
\item Its parent, if it has one, is not the root of a complete subtree.
\end{itemize}
\end{description}

The $\Strip$ operation works as follows: 
 If  $T$ is a complete tree, then return $T$ itself. Note that the root of the $T$ is the only primary root in this case. If $T$ is not a complete tree, then $F$ is obtained as follows: Starting from the root of $T$, traverse the  direct path towards the rightmost leaf of $T$. Remove a node if it is not a primary root. Stop when a primary root or a leaf node (which is a primary root too) is discovered. In figure \ref{sfig: haft-as-join} the $\Strip$ operation removes the nodes indicated by the square boxes. \\
We now give intuition as to why the Strip operation works.
\begin{lemma}
The Strip operation returns the subtrees rooted at all primary roots in the input $\haft$.
\end{lemma}
\begin{proof}
By the definitions of  $\haft$ and primary root, if a vertex is not the root of a complete subtree, its left
child is guaranteed to be a primary root. Thus, either the root of the $\haft$ is a primary root or its left child is. If
the left child is a primary root, there can be no other primary root in the left subtree, so we we return the tree rooted at that
child.  Recursively applying the same test to the right child, we get all the primary roots. 
\end{proof}

\begin{figure}[h!]
\centering
\includegraphics[scale=0.7]{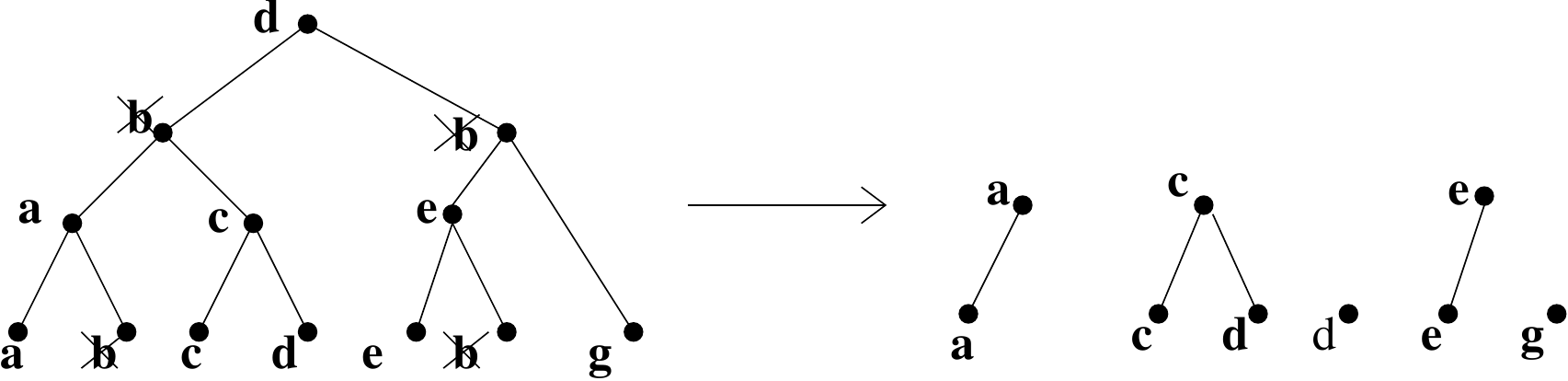}
\caption{Deletion of a node and its helper nodes lead to breakup of RT into components. The Strip operation or a simple variant (for non-hafts) returns a set of complete trees, which can then be merged.} 
\label{fig: HaftStripMerge}
\end{figure}

\subsubsection{Merge}
\label{subsec: haftmerge}
 Every $\haft$ can be represented as a binary number (by lemma~\ref{lemma: hftproperties}).  Merging $\haft$s
is analogous to binary addition of the  binary number representations of these trees. The new binary number obtained is
the representation of the half-full tree corresponding to the merge. This is illustrated in figure~\ref{fig:
haftmergebinary}.\\
 The first step of the $\Merge$ operation is to apply the $\Strip$  operation on the input trees. This gives a forest of
complete trees.  These complete trees can be recombined with the help of extra nodes to obtain a new $\haft$. 
Let $Size(X)$ be the number of nodes in a tree $X$. Consider two complete trees
$T_1$ and $T_2$ (Size($T_1) > Size(T_2$)), with roots $r_{1}$ and $r_{2}$ respectively, and an extra node $v$. To merge
these trees, make $r_{1}$ the left child and $r_{2}$ the right child of $v$ by adding edges between them. The merged
tree is always a $\haft$.
 Thus, the merge operation $\Merge(\haft_1,\haft_2, \ldots)$ is as follows:\\

\begin{enumerate}
\item Apply $\Strip$ to all the hafts to get a forest of  complete trees.
  \item Let $T_{1} , T_{2}, \ldots, T_{k}$ be the $k$ complete trees sorted in ascending order of their
size. Traverse the list from the left, let $T_{i}$ and $T_{i+1}$ be the first two adjacent trees of the same size and $v$ be a
single isolated vertex, join $T_{i}$ and $T_{i+1}$ by making $v$ the parent of the root of $T_{i}$ and the root of
$T_{i+1}$, to give a new tree. Reinsert this tree in the correct place in the sorted list. Continue traversal of the
list from the position of the last merge, joining pairs of trees of equal sizes. At the end of this traversal, we are
left with a sorted list of complete trees, all of different sizes.
\item Let $T_1, T_2, \ldots,  T_l$   be the sorted list of complete trees obtained after the previous step. Traverse
the list from left to right, joining adjacent trees using single isolated vertices.  Let $w$ be a single isolated
vertex. Join $T_1$ and $T_2$ by making the root of $T_2$ the left child and the root of $T_{1}$ the right child of
$w$, respectively. This gives a new haft. Join this haft and $T_3$  by using another available isolated
vertex, making the larger tree ($T_{3}$)  its left child. Continue this process till there is a single haft.
\end{enumerate}

\begin{figure}[h!]
\centering
\includegraphics[scale=0.7]{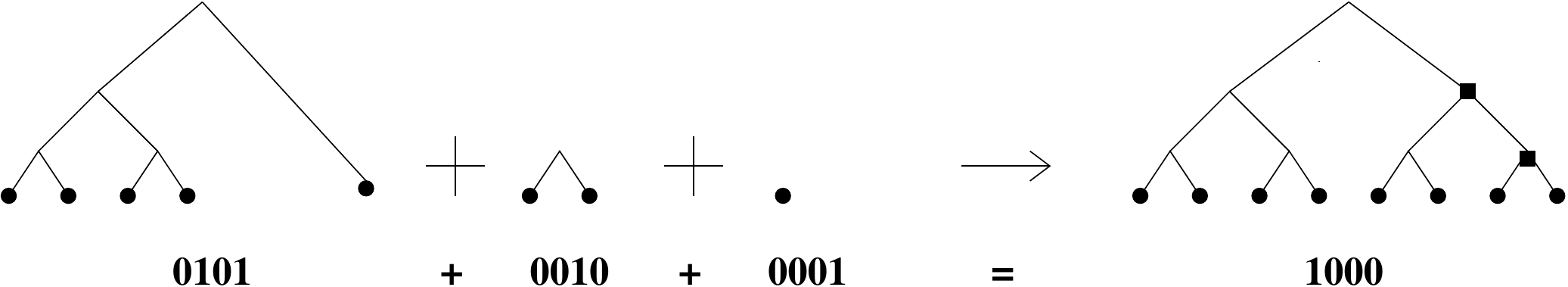}
\caption{Merging three hafts. The square shaped vertices are the isolated vertices used to join complete trees.
Merging is analagous to binary number addition, where the number of leaves are represented as binary numbers.}
\label{fig: haftmergebinary}
\end{figure}

\subsection{Detailed description} 
\begin{figure}[h!]
\centering
\includegraphics[scale = 0.9]{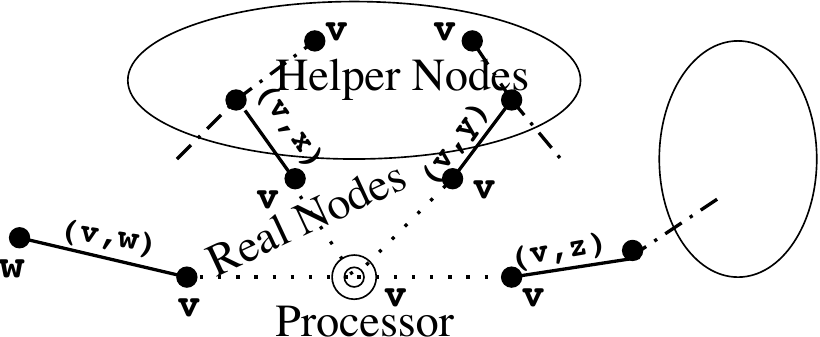}
\caption{The Nodes corresponding to the processor $v$ in the graph $G$. An ellipse denotes a $\RT$ created on deletion of a
neighbor of $v$.}
\label{fig: processornodes}
\end{figure}

As mentioned earlier, deletion of  a node $v$ leads to it being replaced by  a Reconstruction Tree ($\RT(v)$, for short)
in $G$ (Refer to Table~\ref{algo:model-2} for definitions). The $\RT$ is a  $\haft$ (discussed in  Section~\ref{sec:
hafts}) having ``virtual'' nodes as internal nodes  and neighbors of $v$ as the leaf nodes. 
The real network is a homomorphic image of  this virtual graph. The nodes in the virtual graph refer to the
corresponding processor in the network, as shown in Figure~\ref{fig: processornodes}.  The nodes in $G$ corresponding to
an edge of $v$ in $G'$ and forming the leaf nodes in any $\RT$ are called real nodes, and those internal to a $\RT$
and simulated by the real nodes (more precisely, by the processor) are called helper nodes. There is one real node and
at most one helper node corresponding to an edge of $v$ in $G'$ i.e. to an edge formed when $v$ or $v$'s neighbor
joined the network.   In Table~\ref{tab: nodedata} we list the information each processor $v$ requires for each edge in
order to execute the ForgivingGraph algorithm. When one of the nodes of the edge gets deleted, in $G$, that node may be
replaced by a helper node.  This end point of the edge is stored in the field  $v.\Endpoint$. For an edge $(v,x)$, if $x$ is a real
node then the field $v.\Endpoint$ is simply the node $x$. If the node $x$ gets deleted, the new endpoint may be a helper
node, though we  still refer to this edge as $(v,x)$ i.e. by its name in $G'$. Moreover, the processor may now simulate
a helper node corresponding to this edge. Since each edge is uniquely identified, the real nodes and helper nodes
corresponding to that edge can also be uniquely identified. This identification is used by the processors to pass
messages along the correct paths. The \textsc{Forgiving graph} algorithm is given in pseudocode form in Algorithm~\ref{algo: forgiving}
alongwith the required subroutines. For ease of description, the real and helper nodes belonging to the same processor
may not be explicitly distinguished in the code.

\begin{table}[h!]
\begin{tabular}{|l|p{4.5in}|}
\hline 
\textbf{Processor v: Edge(v,x)}&\\ \cline{2-2}
\hline 
\textbf{Real node fields}&\\ \cline{2-2}
\hline
  \texttt{Endpoint}& The node that represents the other end of the edge. For edge(v,x) this will be node $x$ if $x$ is
alive or $\RTparent$ if $x$ is not. \\
  \texttt{hashelper}& (boolean field). True if there is a helper node simulated by $v$ corresponding to this edge.\\
  \texttt{RTparent} & Parent of $v$ in $\RT$. Non NULL only if $x$ has been deleted.\\
   \texttt{Representative}& This is $v$ itself. Field used during merging of $\RT$s. \\
\hline
\textbf{Helper node fields} & Fields for helper node corresponding to the edge. Non NULL only if the helper node exists.
Sometimes, we will refer to a helper field as \emph{edge.helper.field}\\
\cline{2-2}
\hline
 \texttt{hparent}& Parent of helper node. \\
 \texttt{hrightchild}& Right Child of helper node. \\
 \texttt{hleftchild}& Left Child of helper node. \\
 \texttt{height} &  Height of the helper node.\\
 \texttt{childrencount} & The number of descendants of the helper node.\\
 \texttt{Representative}& The unique leaf node of a subtree of a $\RT$ that does not have a helper node in that
subtree. This node is used during merging of $\RT$s.\\
\hline
\end{tabular}
\caption{The fields maintained by a processor $v$ for edge$(v,x)$, which is an edge in $G'$, the graph of only original nodes and  insertions.}
\label{tab: nodedata}
\end{table}

\begin{figure}[h!]
\centering
\includegraphics[scale=0.35]{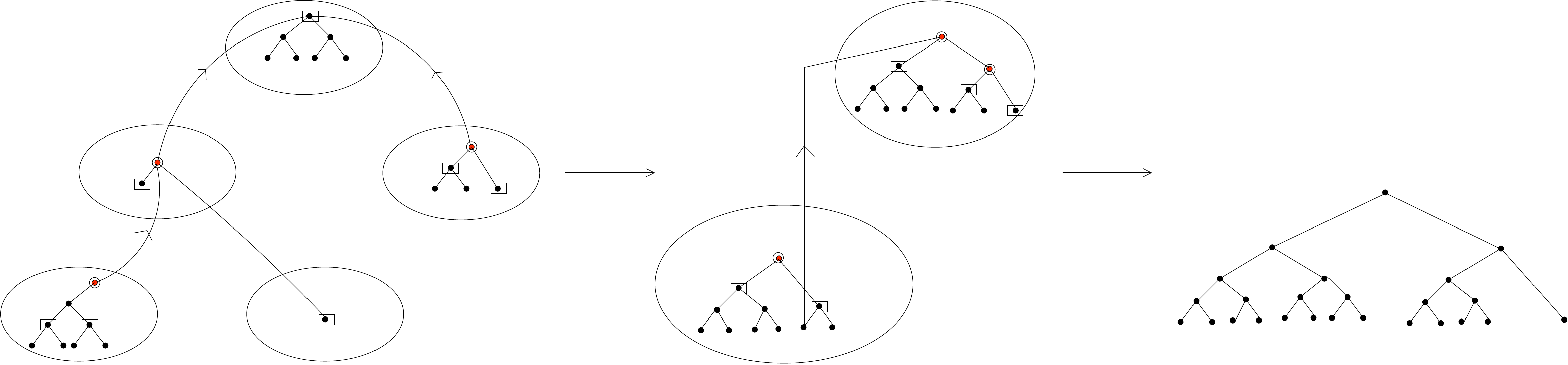}
\caption{On deletion of a node $v$, The RTs to be merged are connected by  $BT_{v}$ which is a binary tree. The RTs merge
from the bottom up with their parents till a single RT is left. The nodes in the square boxes are the primary roots. The (red
color) nodes in the circle are excess nodes removed at each step.}
\label{fig: Anchormerge}
\end{figure}

\begin{figure}[h!]
\centering
\includegraphics[scale=0.7]{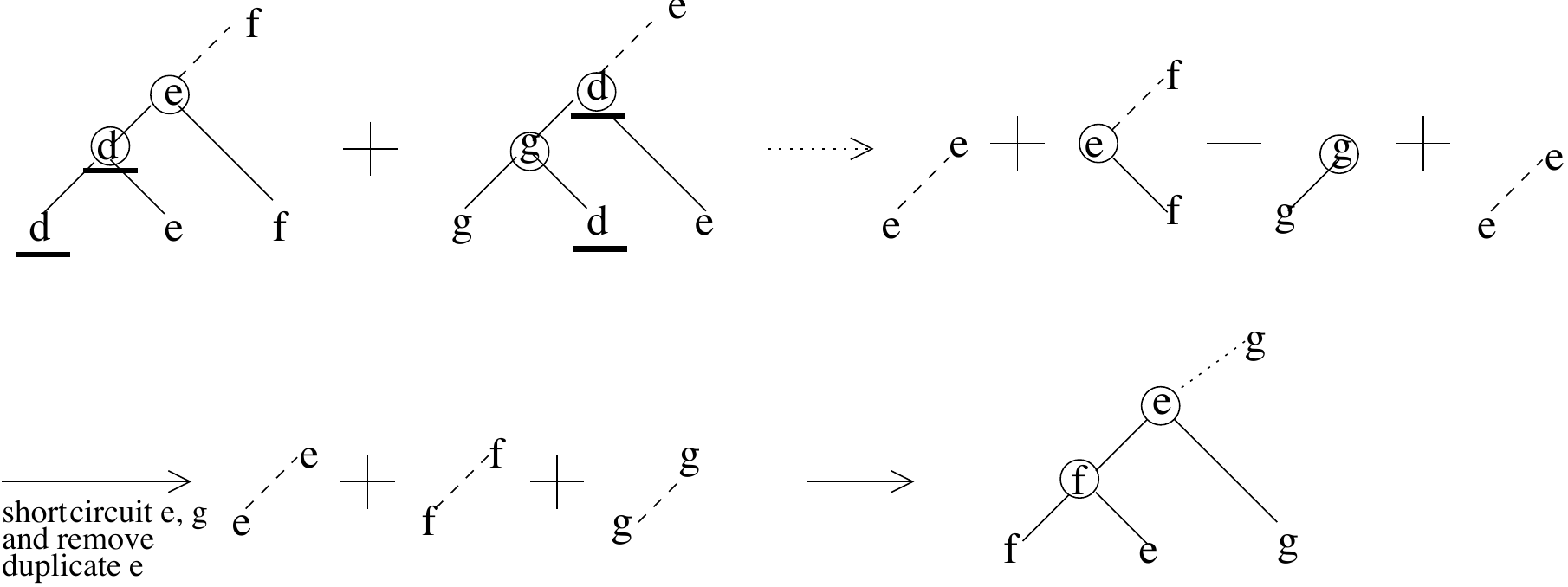}
\caption{The underlined node $d$ and corresponding helpers are deleted. This leads to the graph breaking into
components which are then merged using $BT_{d}$ (the binary tree of anchors) and the primary roots in the components. The dashed edges
show the representative for that node.}
\label{fig: RTmerge}
\end{figure}


On deletion of a node, the repair proceeds in two phases. The first phase is a quick $O(1)$ phase in which the neighbors
of the deleted node connect themselves in the form of a binary tree (Algorithm~\ref{algo: fixnode}). These neighbors represent the
independent components created on deletion of the node. Some of these components may not be hafts. We shall refer to
such a subtrees as a $\RTfragment$. Let $v$ be the processor deleted. Then, we call this tree formed by the $v$'s
neighbors as $BT_{v}$ and the nodes forming $BT_{v}$ as \emph{anchors}. Formally, we define an anchor as follows:
\begin{description}
\item[Anchor]: An anchor is the unique designated node in a $\RT$ or $\RTfragment$ that takes part in the binary
tree $BT_{v}$.
\end{description}
 In phase 2,  the $\RT$s and $\RTfragment$s forming $BT_{v}$ have to be merged (Figure~\ref{fig: Anchormerge}). We are only interested in
the complete trees in these since we can discard all other helper nodes. The anchors send probe messages to discover
the primary roots which head these complete trees (Algorithm~\ref{algoline: findprroots}). This is similar to the Strip
operation described in Section~{\ref{subsec: haftstrip}}. The
nodes maintain information about their height and number of their children in their $\RT$ or $\RTfragment$. Thus,
they are able to identify themselves as primary roots. At the same time, the nodes outside the complete trees are
identified and marked for removal. The complete trees are then merged pairwise in a bottomup fashion till only a single
haft remains. This is
illustrated in figure \ref{fig: Anchormerge}. At each round, every leaf $\RT$ in $BT_{v}$ will merge with its parent
$\RT$. This can be done in parallel, so that the  number of rounds of merges will be equivalent to the height of the
tree. For two trees to merge, as shown in the Merge operation (Section~\ref{subsec: haftmerge}), an additional node is
needed that will become the parent of these two trees. This node must be simulated by a real node that is not already
simulating a helper node in the trees. Since the number of internal  nodes in a tree is one less than the leaf nodes,
there is exactly one such leaf node for each tree. The roots of these two trees keep the identity of this node. This is
stored in the field Representative (Table~\ref{tab: nodedata}). More  formally, we define a representative as follows:

\begin{description}
\item[Representative]
Given a node $y$, the representative of $y$ is a real node, decided as follows:
\begin{itemize}
 \item If $y$ is a real node, then $y$ itself.
 \item If $y$ is a helper node, then the unique leaf node of $y$'s subtree in $y$'s $\RT$ that does not have a helper node in that
subtree. 
\end{itemize}
\end{description}

 We now describe a mechanism for merging that we call the representative mechanism. Each node has a representative defined earlier. When
two trees (Note that a tree may even be a single node) are merged (Algorithm~\ref{algo: makeRT} and Algorithm~\ref{algo:
ComputeHaft} ), the representative of
the root of the bigger tree (or of one of the trees, if they have the same size) instantiates a new helper node, and
makes the two roots its children. The new helper node will now inherit as its representative the representative of the
root of the other tree, since this is the node in the merged tree that does not have a helper node in the tree. An
example of merging using this mechanism is shown in Figure~\ref{fig: RTmerge}.
 At the end of each round, we have a  new set of leaf $\RT$s. Each new leaf is now a
merged haft of the previous leaves and their parent. We need a new anchor for this haft. We can continue having the
anchor of the parent $\RT$ or $\RTfragment$ as the anchor. However, this
 node may be one of the extra nodes marked for removal. In this case, the anchor designates one of the nodes that was
 a primary root in its $\RT$ as the new anchor, passes on its links and removes itself.
 Now, the newly formed leaf hafts may have primary roots which are different from those of the previous ones. The new
anchor will again send probe messages and gather this information and inform the new primary roots of their role. This
process will continue till we are left with a single $\RT$. This is shown in Figure~\ref{fig: Anchormerge}. 

%
\section{Results}
\label{sec: Results}

\subsection{Upper Bounds}
\label{subsec: upperbounds}
Let $G'$ be the graph consisting solely of the original nodes and insertions without regard to deletions and
healings. Let $G_{T}$ and $G'_{T}$ be the graphs at time $T$. 

\begin{lemma}
\label{lemma: hnodecounts}
Given the real node $v$ in $G$ corresponding to an edge $(v,x)$ in $G'$,
\begin{enumerate}
\item \label{lmpart: onehelper} There can be at most one helper node in $G$ corresponding to $v$.
\item  \label{lmpart: twohelper} During the Repair phase, there can be at most two helper nodes corresponding to the edge $(v,x)$. Moreover, one of these could also be an anchor in $BT_{v}$
\end{enumerate}
\begin{proof}
 As stated earlier, there is only one real node in $G$ corresponding to an edge in $G'$ (Figure~\ref{fig:
processornodes}). Also, any real node can only form a leaf node of a $\RT$, and a helper node can only be an
internal node. 
We prove  part~\ref{lmpart: onehelper} by contradiction. Suppose there are two helper nodes in $G$ corresponding to
the real node $v$. Let us call these nodes $v'$ and $v''$. The following cases arise:
\begin{enumerate}
\item \emph{$v'$ and $v''$ belong to different $\RT$s:}\\
 By the representative mechanism, a helper node is created only if the real node that simulates it is the representative
of a node (e.g. in line~\ref{algocode: mergeeqrep} in Algorithm~\ref{algo: ComputeHaft}). By definition, the
representative of a node is a unique leaf node in the subtree headed by that node in its $\RT$. If both $v'$ and $v''$
exist and  belong to different $\RT$s, this implies that node $v$ exists as a leaf node in two different $\RT$s. This is
a contradiction.
\item \emph{$v'$ and $v''$ belong to the same $\RT$:}\\
 Without loss of generality, assume that the  $v''. height \ge v'.height$. The following cases arise:
 \begin{enumerate}
  \item \emph{$v'$ is a node in the subtree headed by $v''$:}
     Note that by the representative mechanism, in a subtree, an internal helper node will be created earlier than the
root of the subtree. Thus, node $v'$ will be created before node $v''$.
Let node $y$ be the child of node $v''$ that had $y.\representative = v''$ when $v''$ was created. However, 
   $y.\representative$ could not have been $v''$, since by definition, $y.\representative$ has to be the unique leaf 
node not  simulating a helper node in $y$'s subtree, but $v$ is already simulating $v'$ in $v''$'s subtree.
  \item \emph{$v'$ is a node not in the subtree headed by $v''$:}
   Again, the representaive mechanism and definition of a representative implies that node $v$ was a
representative in two non-intersecting subtrees in the same $\RT$. This implies that node $v$ occurs as a leaf
twice in that $\RT$. This is not possible. 
 \end{enumerate}
\end{enumerate}
 Now, we prove part~ \ref{lmpart: twohelper}. As stated earlier, at each stage of the merge procedure, leaf $\RT$s or
$\RTfragment$s in $BT_{v}$ will merge with their parent. Suppose that $v'$ is a helper node simulated by real node $v$,
and $v'$ is not part of any complete subtree in such a $\RTfragment$ or $\RT$. This means that $v'$ will be marked red
and removed when this stage of merge is completed (Refer Figure~\ref{fig: Anchormerge}). Let node $y$ be the root of the
 complete subtree (i.e. a primary root in that $\RTfragment$) that has $v$ as a leaf node. 
Node $v'$ is an ancestor of node $y$ since $v'$ cannot a descendant. By definition, $y.\representative = v$, since $v$
will be the unique leaf node in $y$'s subtree not simulating a helper node in that subtree.
 When the trees are being merged, $v$ may be asked to create another helper node. Thus, $v$ may have two helper nodes.
 Also, each $\RT$ or $\RTfragment$ has exactly one anchor node. This anchor may be $v'$ or another node. Thus, in the
repair phase, a real node may simulate upto two helper nodes, and one of these helper nodes may be an anchor.
 However, node $v'$ will be removed as soon as this stage is completed, and if $v'$ was an anchor, a new anchor is
chosen from the existing nodes. Since at the end of the merge, $BT_{v}$  collapses to leave one $\RT$, the extra helper
nodes and the edges from the anchor nodes are not present in $G$, thus, not contradicting part~\ref{lmpart: onehelper}.

\end{proof}
\end{lemma}

\begin{lemma}
\label{lemma: cost}
After each deletion, the repair can take $O(\log d\log n)$ time to exchange $O(d \log n)$ messages of size $O(log
n)$, where $d$ is the degree of the deleted node.
\end{lemma}

\begin{proof} There are mainly two types of messages exchanged by the algorithm. They are the probe messages sent by the
\textsc{FindPrRoots()} (Algorithm~\ref{algo: findprroots}) within a $\RT$ and the messages containing the information
about the primary roots exchanged by the anchors in $BT_{v}$ and among the primary roots themselves (Algorithm~\ref{algo:
haftmerge}: \textsc{ComputeHaft()}). Let $\size(BT_{v})$ be the number of $\RT$s of $BT_{v}$. Since a helper
node can split a $\RT$ into maximum 3 parts, and there can be at most $d$ helper nodes, where $d$ is the degree of the
deleted node $v$, $\size(BT_{v}) = 3d$.
Now, let us calculate the number of messages:
\begin{itemize}
\item \emph{Probe messages (Algorithm~\ref{algo: findprroots})}: A probe message is generated by a
 an anchor of a $\RT$. This is similiar to the \emph{Strip} operation
(Section~\ref{subsec: haftstrip}). The path that the probe message follows is the direct path from the originating
node to the rightmost node of the $\RT$. At the most 2 messages can be generated for every node on the way. Further,
there can be one confirmatory message transmitted from the primary roots back to the anchor. Let $numnodes$ be 
the number of nodes and $numprobes$ be number of probe messages sent in a single $\RT$.
Thus, 
\begin{eqnarray*}
 numprobes & \le & 3 \log numnodes  \\
   & \le & 3 \log n  
\end{eqnarray*}
\item \emph{Exchange of primary roots lists (Algorithm~\ref{algo: haftmerge})}:  At each step of Algorithm~\ref{algo:
bottomupmerge} (\textsc{BottomupRTMerge()}), leaves in $BT_{v}$ merge with their parents. Let $rtlistmsgs$
be the number of messages exchanged for every such merge.  The anchors of the leaves of $BT_{v}$ send their primary
roots lists to the parent, which in turn can send both it's list and the sibling's list to the child.  Thus,
$rtlistmsgs = 4$. In addition, every anchor will send this list to the primary roots in its $\RT$, generating at most
another $\log n$ messages (Let us call this $AtoRmsgs$).
\end{itemize}
As stated earlier, in the $BT_{v}$, leaves merge with their parents. The number of such merges
before we are left with a single $\RT$ is $\lceil \size(BT_{v})/2 - 1 \rceil$. Also, at most 3 $\RT$s are involved in
each merge. Let $totmessages$ be the total number of messages exchanged. Hence,

\begin{eqnarray*}
totmessages & = & \lceil \size(BT_{v})/2 - 1 \rceil( 3 ( numprobes  + AtoRmsgs) + rtlistmsgs )\\
 & \le & \lceil 3d/2 - 1 \rceil ( 12 \log n + 4 )\\
 & \in & O(d \log n)
\end{eqnarray*}
 In $BT_{v}$, leaves and their parents merge. This can be done in parallel such that each time the level of $BT_{v}$
reduces by one. Within each $\RT$, the time taken for message passing is still bounded by $O(\log n)$
assuming constant time to pass a message along an edge. Since there are at most $\lceil log d \rceil$ levels, the time
taken for passing the messages is $O(\log d \log n)$. The biggest message exchanged may have constant size information
about the primary roots of upto two $\RT$s. This may be the message sent by a parent $\RT$ in $BT_{v}$ to its children
$\RT$. Since there can be at most $O(\log n)$ primary roots, the size of messages is $O(\log n)$.
\end{proof}

Here, we state our main theorem.

\begin{theorem}
 
The Forgiving Graph has the following properties:
\label{theorem: forgiving}
\begin{enumerate}
\item\label{th: degree} 
 \emph{Degree increase:} For any node $v$, $d(v,G_{T}) \le 3 \times d(v,G'_{T})$, where $d$ is the degree of the node $v$. 
\item\label{th: stretch} 
\emph{Stretch:} For any pair of nodes $x$ and $y$, $distance(x,y, G_{T}) \le (\log n) \times distance(x,y,G'_{T})$.
\item \label{th: cost}  \emph{Cost:} After each deletion, the repair can take upto $O(\log d\log n)$ time with $O(d \log
n)$ messages of size upto $O(log n)$, where $d$ is the degree of the deleted node.
\end{enumerate}
\end{theorem}

\begin{proof}
 Part~\ref{th: degree} follow directly by construction of our algorithm.  For part~\ref{th: degree}, we note that for a
node $v$, any degree increase for $v$ is imposed by the  edges of its helper node to $\hparent$($v$) and $\hchildren(v)$. From lemma~\ref{lemma: hnodecounts} part~\ref{lmpart:
onehelper}, we know that, in $G$, node $v$ can play the role of at most one helper node for any of its neighbors in
$G'$ at  any time (i.e. $d(v,G'_{T})$ ).
The number of $\hchildren$ of a helper node are never more than $2$, because the reconstruction trees are binary trees. 
Thus the total degree of $v$ ( $d(v,G_{T})$) is at most $3$ times its degree in $G'$ ($d(v,G'_{T})$). 


We next show Part~\ref{th: stretch}, that the stretch of the Forgiving Graph is  $O(D \log n)$, where $n$ is the number
of nodes in $G_{T}$. The distance between any two nodes $x$ and $y$ cannot increase by more than the factor of the
longest path in the largest $\RT$ on the path between $x$ and $y$. This factor is $\log n$ at the maximum.

 Part~\ref{th: cost} follows from Lemma~\ref{lemma: cost}. Note that besides the commuication of the messsages
discussed, the other operations can be done in constant time in our algorithm.
 
\end{proof}

\pagebreak

\subsection{Lower Bounds}

\label{subsec: lowerbounds}

\begin{theorem}
Consider any self-healing algorithm that ensures that: 1) each node increases its degree by a multiplicative factor of 
at most $\alpha$, where $\alpha \geq 3$; and 2) the stretch of the graph
increases by a multiplicative factor of at most $\beta$. Then, for some initial graph with $n$ nodes, it must be the case that
 $\beta \geq \frac{1}{2} \log_{\alpha-1}( n - 1)$.
\end{theorem}

\begin{proof}
Let $G$ be a star on $n$ vertices, where $x$ is the root node, and $x$ has an edge with each of the
other nodes in the graph. The other nodes (besides $x$) have a degree of only 1. Let $G'$ be the graph created after the
adversary deletes the node $x$.  Consider a breadth
first search tree, $T$, rooted at some arbitrary node $y$ in $G'$.  We know that the self-healing algorithm can increase
the degree of each node by at most a factor of $\alpha$, thus every node in $T$ besides $y$ can have at most $\alpha-1$
children. 
Let $h$ be the height of $T$.  Then we know that $1 + \alpha \sum_{i=0}^{h-1}
(\alpha-1)^{i} \geq n-1$.  This implies that  $(\alpha-1)^{h} \geq n-1$ for $\alpha \geq 3$, or ${h} \geq
\log_{\alpha-1}(n-1)$.  Let $z$ be a leaf node in $T$ of largest depth. Then, the distance between $y$ and $z$ in $G'$ is $h$ and the distance 
between $y$ and $z$ in $G$ is 2. Thus, $\beta \ge h/2$, and  $2\beta 
\geq \log_{\alpha-1} (n-1)$, or $\beta \geq \frac{1}{2}\log_{\alpha-1}( n-1)$. 
\end{proof}
\noindent
We note that this lower-bound compares favorably with the general
result achieved with our data structure.  

\section{Conclusion}
We have presented a distributed data structure that withstands
repeated adversarial node deletions by adding a small number of new
edges after each deletion.  Our data structure is efficient and ensures two key
properties, even in the face of both adversarial deletions and adversarial insertions.
First, the distance between any pair of nodes never increases by more than a $\log n$ multiplicative factor than what the distance would be without the adversarial deletions.  Second, the degree of any node never increases by more than a $3$ multiplicative factor.

Several open problems remain including the following. Can we design algorithms
for less flexible networks such as sensor networks?  For example, what
if the only edges we can add are those that span a small distance in
the original network?  Can we extend the concept of
self-healing to other objects besides graphs?  For example, can we
design algorithms to rewire a circuit so that it maintains its
functionality even when multiple gates fail?

\pagebreak
\newpage
\bibliography{selfheal} 

\begin{thebibliography}{10}

\bibitem{anderson01RON}
David Andersen, Hari Balakrishnan, Frans Kaashoek, and Robert Morris.
\newblock Resilient overlay networks.
\newblock {\em SIGOPS Oper. Syst. Rev.}, 35(5):131--145, 2001.

\bibitem{BomanSAS06}
Iching Boman, Jared Saia, Chaouki~T. Abdallah, and Edl Schamiloglu.
\newblock Brief announcement: Self-healing algorithms for reconfigurable
  networks.
\newblock In {\em Symposium on Stabilization, Safety, and Security of
  Distributed Systems(SSS)}, 2006.

\bibitem{doverspike94capacity}
Robert~D. Doverspike and Brian Wilson.
\newblock Comparison of capacity efficiency of dcs network restoration routing
  techniques.
\newblock {\em J. Network Syst. Manage.}, 2(2), 1994.

\bibitem{frisanco97capacity}
T.~Frisanco.
\newblock Optimal spare capacity design for various protection switchingmethods
  in atm networks.
\newblock In {\em Communications, 1997. ICC 97 Montreal, 'Towards the Knowledge
  Millennium'. 1997 IEEE International Conference on}, volume~1, pages
  293--298, 1997.

\bibitem{goel04resilient}
Sanjay Goel, Salvatore Belardo, and Laura Iwan.
\newblock A resilient network that can operate under duress: To support
  communication between government agencies during crisis situations.
\newblock {\em Proceedings of the 37th Hawaii International Conference on
  System Sciences}, 0-7695-2056-1/04:1--11, 2004.

\bibitem{hayashi2005}
Yukio Hayashi and Toshiyuki Miyazaki.
\newblock Emergent rewirings for cascades on correlated networks.
\newblock cond-mat/0503615, 2005.

\bibitem{HayesPODC08}
Tom Hayes, Navin Rustagi, Jared Saia, and Amitabh Trehan.
\newblock The forgiving tree: a self-healing distributed data structure.
\newblock In {\em PODC '08: Proceedings of the twenty-seventh ACM symposium on
  Principles of distributed computing}, pages 203--212, New York, NY, USA,
  2008. ACM.

\bibitem{holme-2002-65}
Petter Holme and Beom~Jun Kim.
\newblock Vertex overload breakdown in evolving networks.
\newblock {\em Physical Review E}, 65:066109, 2002.

\bibitem{iraschko98capacity}
Rainer~R. Iraschko, M.~H. MacGregor, and Wayne~D. Grover.
\newblock Optimal capacity placement for path restoration in stm or atm
  mesh-survivable networks.
\newblock {\em IEEE/ACM Trans. Netw.}, 6(3):325--336, 1998.

\bibitem{medard99redundant}
Muriel Medard, Steven~G. Finn, and Richard~A. Barry.
\newblock Redundant trees for preplanned recovery in arbitrary vertex-redundant
  or edge-redundant graphs.
\newblock {\em IEEE/ACM Transactions on Networking}, 7(5):641--652, 1999.

\bibitem{motter-2004-93}
Adilson~E Motter.
\newblock Cascade control and defense in complex networks.
\newblock {\em Physical Review Letters}, 93:098701, 2004.

\bibitem{motter-2002-66}
Adilson~E Motter and Ying-Cheng Lai.
\newblock Cascade-based attacks on complex networks.
\newblock {\em Physical Review E}, 66:065102, 2002.

\bibitem{murakami97comparative}
Kazutaka Murakami and Hyong~S. Kim.
\newblock Comparative study on restoration schemes of survivable {ATM}
  networks.
\newblock In {\em {INFOCOM} (1)}, pages 345--352, 1997.

\bibitem{SaiaTrehanIPDPS08}
Jared Saia and Amitabh Trehan.
\newblock Picking up the pieces: Self-healing in reconfigurable networks.
\newblock In {\em IEEE International Parallel \& Distributed Processing
  Symposium}, 2008.

\bibitem{caenegem97capacity}
B.~van Caenegem, N.~Wauters, and P.~Demeester.
\newblock Spare capacity assignment for different restoration strategies in
  mesh survivable networks.
\newblock In {\em Communications, 1997. ICC 97 Montreal, 'Towards the Knowledge
  Millennium'. 1997 IEEE International Conference on}, volume~1, pages
  288--292, 1997.

\bibitem{xiong99restore}
Yijun Xiong and Lorne~G. Mason.
\newblock Restoration strategies and spare capacity requirements in
  self-healing atm networks.
\newblock {\em IEEE/ACM Trans. Netw.}, 7(1):98--110, 1999.

\end{thebibliography}
\bibliographystyle{plain}

\pagebreak
\appendix

\section{ForgivingGraph PseudoCode}

\floatname{algorithm}{Algorithm}
\begin{algorithm}[h!]
\begin{algorithmic}[1]
\STATE Given a Graph $G(V,E)$
\REQUIRE{each node of G has a unique ID}
\FOR{each node  $v\in G$} 
\STATE \textsc{Init(v)}.
\ENDFOR
\WHILE {true}
\IF{a vertex $v$ is inserted}
\STATE vertex $v$ and new neighbors add appropriate edges.
\STATE \textsc{Init(v)}.
\ELSIF{a vertex $v$ is deleted}
\STATE \textsc{DeleteFix(v)}
\ENDIF
\ENDWHILE
\end{algorithmic}
\caption{\textsc{Forgiving graph}: The main function.}
\label{algo: forgiving}
\end{algorithm}

 \begin{algorithm}[h!]
 \caption{\textsc{Init(v)}: initialization of the node $v$} 
\label{algo: init}
\begin{algorithmic}[1]
\FOR{each $\edge (v,x)$}
\STATE $(v,x).\representative = v$
\STATE set other fields to NULL.
\ENDFOR
\end{algorithmic}
\end{algorithm}

\begin{algorithm}[h!]
\caption{\textsc{DeleteFix($v$)}: Self-healing on deletion of a node }
\label{algo: fixnode}
\begin{algorithmic}[1]
\STATE $\Nset = \{ \}$
\FOR{each $\edge (v,x)$}
\IF{$(v,x).hashelper = \True $}
\STATE $\Nset = \Nset \cup (v,x).\hparent \cup (v,x).\hrightchild$
\ENDIF
\STATE $\Nset = \Nset \cup  (v,x).\Endpoint$
\ENDFOR
\STATE \label{algoline: BTquickfix} The nodes in $\Nset$ make new edges to make a balanced binary tree $\BT_{v} (\Nset,E_{v})$.
\STATE \textsc{BottomupRTMerge($\BT_{v},v$)}
\STATE delete the edges $E_{v}$.
\end{algorithmic}
\end{algorithm}

\begin{algorithm}[ph!]
\caption{\textsc{BottomupRTMerge($\BT_{v}, v$)}: The nodes of $\BT_{v}$ merge  their $\RT$s starting from the leaves going
up forming a new $\BT_{v}$. }
\label{algo: bottomupmerge}
\begin{algorithmic}[1]
\IF{$\BT_{v}$ has only one node}
\STATE return
\ENDIF
\FOR{$y \in \BT_{v}$}\label{algoline: findprroots}
 \IF{$y$ is a real node}
 \STATE Let $\PrRoots(y) \leftarrow y$
 \ELSIF{$y = (v,x).\Endpoint$}
 \STATE \textsc{FindPrRoots($y, 1,\real(v), \True$ )}
 \ELSIF {$\helper(y).hparent = v$ OR $\helper(y).hleftchild = v$ OR $\helper(y).hrightchild = v$}
  \STATE Let $\PrRoots(y) \leftarrow$ \textsc{FindPrRoots($y, v.\childrencount,\helper(v),
\True$)}
  \ELSE
  \STATE  Let $\PrRoots(y) \leftarrow$ \textsc{FindPrRoots($y, v.\childrencount,\helper(v), \False$)}
  \ENDIF
\ENDFOR
\FOR{all nodes $y$ s.t. node $y$ is a parent of a leaf in $\BT_{v}$}
\IF{$y$ has two children in $BT_{v}$}
\STATE \textsc{Haft\_Merge}($y$, $y$ 's left child in $BT_{v}$, $y$ 's right child in $BT_{v}$)
\ELSE
\STATE \textsc{Haft\_Merge}($y$, $y$'s left child, NULL)
\ENDIF
\ENDFOR
\STATE \textsc{BottomupRTMerge($\BT_{v}$)} \COMMENT{The new leaf nodes merge again till only one is left.}
\end{algorithmic}
\end{algorithm}


\begin{algorithm}[ph!]
\caption{\textsc{FindPrRoots}($y$, numchild, sender, Breakflag): Find the primary roots in the $\RT$ beginning
with node $y$. If Breakflag is set the tree is  a component of the $\RT$ formed due to the deletion.}
\label{algo: findprroots}
\begin{algorithmic}[1]
\IF{Breakflag = $\True$ AND (sender = $y.\hrightchild$ OR sender = $y.\hleftchild$ )}
  \STATE $y.\childrencount = y.\childrencount$ - numchild
\ENDIF
\IF{$y.\childrencount = 2^{y.\height}$} 
  \IF {\textsc{TestPrimaryRoot($y$)} = $\True$} 
  \STATE return \{$y$,\textsc{FindPrRoots($y.\hparent, 0, y$, Breakflag)} \}
  \ELSE
  \STATE return \{\textsc{FindPrRoots($y.\hparent, 0, y$, Breakflag)} \} \COMMENT{Node itself not a primary
root but parent maybe.}
  \ENDIF
\ELSE
\STATE mark node red
  \IF{exists($y.\hleftchild$) AND sender != $y.\hleftchild$}
   \STATE \textsc{FindPrRoots($y.\hleftchild, y.\childrencount, y$, Breakflag)}
   \ELSIF{exists($y.\hrightchild$) AND sender != $y.\hrightchild$}
   \STATE \textsc{FindPrRoots($y.\hrightchild, y.\childrencount, y$, Breakflag)}
   \ELSIF{exists($y.\hparent$) AND sender != $y.\hparent$}
   \STATE \textsc{FindPrRoots($y.\hparent, y.\childrencount, y$, Breakflag)}
  \ENDIF
\ENDIF
\end{algorithmic}
\end{algorithm}


\begin{algorithm}[ph!]
\caption{\textsc{TestPrimaryRoot($y$)}: Tell if helper node $y$ is a primary root in $\RT$ }
\label{algo: testprimaryroot}
\begin{algorithmic}[1]
\IF{$y.\childrencount = 2^{y.\height}$} 
 \IF{$y.\hparent = NULL$}
\STATE return $\True$
\ELSIF{$y.\hparent.\childrencount \neq  2^{y.\hparent.\height}$}
\STATE return $\True$
 \ENDIF
 \ENDIF
\STATE return $\False$ 
\end{algorithmic}
\end{algorithm}

\begin{algorithm}[ph!]
\caption{\textsc{Haft\_Merge($p,l,r$)}: Merge the hafts mediated by anchors $p,l$ and $r$}
\label{algo: haftmerge}
\begin{algorithmic}[1]
\STATE Nodes $p, l$ and $r$ exchange $\PrRoots(p), \PrRoots(l), \PrRoots(r)$
\STATE Let $\RT \leftarrow $ \textsc{MakeRT($\PrRoots(p), \PrRoots(l), \PrRoots(r)$)}
  \IF{$p$ is marked red} 
  \STATE $p$ transfers its edges in $BT_{v}$ to one of $\PrRoots(p)$ \COMMENT{$p$ needs to be removed, $BT_{v}$ needs
to be maintained}
  \ENDIF
 \STATE Remove all helper nodes marked red \COMMENT{Some helper nodes marked red may have been reused and unmarked by
\textsc{MakeRT}}
\end{algorithmic}
\end{algorithm}

\begin{algorithm}[ph!]
\caption{\textsc{MakeRT}(PRoots1, PRoots2, PRoots3): The sets of Primary roots make a new RT }
\label{algo: makeRT}
\begin{algorithmic}[1]
\FOR{all $y \in (\mathrm{PRoots1} \cup \mathrm{PRoots2} \cup \mathrm{PRoots3}) $ }
  \STATE Let $\Haftmergeprint \leftarrow$ \textsc{ComputeHaft}(PRoots1, PRoots2, PRoots3)
  \STATE Make helper nodes and set fields and make edges according to  $\Haftmergeprint$
 \ENDFOR
\end{algorithmic}
\end{algorithm}

\begin{algorithm}[ph!]
\caption{\textsc{ComputeHaft($\mathrm{PRoots1, PRoots2, PRoots3})$}: (Implementation of Haft Merge) The primary roots
compute the new haft}
\label{algo: ComputeHaft}
\begin{algorithmic}[1]
\STATE Let $R =  \mathrm{PRoots1} \cup \mathrm{PRoots2} \cup \mathrm{PRoots3} $
\STATE Let $L =  R$ sorted in ascending order of number of children, NodeID
\STATE Suppose $L$ is $(r_{1}, r_{2}, \ldots, r_{k})$ where the $r_{i}$ are the $k$ ordered primary roots.
\STATE set $ctr = 1, count = k$
\WHILE{$ctr < count$}
  \IF{$r_{ctr}.\numchildren = r_{ctr+1}.\numchildren$ }
  \STATE \label{algocode: mergeeqrep} Make helper node $\helper(r_{ctr}.\representative)$. Initialise all its fields
to NULL.
  \STATE Make  $\helper(r_{ctr}.\representative)$ the parent of $r_{ctr}$ and $r_{ctr + 1}$
  \IF{$r_{ctr}$ is a real node}
   \STATE  Set $\helper(r_{ctr}.\representative).\height = 1$ 
   \ELSE
    \STATE Set $\helper(r_{ctr}.\representative).\height = 2 r_{ctr}.height$ 
  \ENDIF
  \STATE Set $\helper(r_{ctr}.\representative).\representative = r_{ctr+1}.\representative$ 
  \STATE remove $r_{ctr}, r_{ctr + 1}$ and insert  $\helper{r_{ctr}\representative}$ in the correct position in $L$.
  \STATE set $ctr \leftarrow ctr - 1$, $count \leftarrow count - 1$
  \ENDIF
  \STATE set  $ctr \leftarrow ctr + 1$,
\ENDWHILE
\STATE set $ctr = 1$
\WHILE{$ctr < count$}
\STATE \label{algocode: mergeneqrep} Make helper node $\helper(r_{ctr+1}.\representative)$. Initialise all its fields to
NULL
\STATE Set  $\helper(r_{ctr + 1}.\representative).\hleftchild = r_{ctr+1}$
\STATE Set  $\helper(r_{ctr + 1}.\representative).\hrightchild = r_{ctr}$
\STATE Set $\helper(r_{ctr + 1}.\representative).\height =  r_{ctr+1}.\height + 1$ 
\STATE Set $\helper(r_{ctr + 1}.\representative).\representative =  r_{ctr}.\representative$ 
\STATE In $L$, replace $r_{ctr + 1}$ by  $\helper(r_{ctr + 1}.\representative)$ 
\ENDWHILE
\end{algorithmic}
\end{algorithm}

\end{document}